\documentclass[fleqn,11pt]{wlscirep}
\usepackage[utf8]{inputenc}
\usepackage[T1]{fontenc}
\usepackage{siunitx}  
\usepackage[version=4]{mhchem}
\usepackage{float}

\usepackage{soul} 

\usepackage{lineno}
\DeclareSIUnit\electron{\mathrm{e^-}}
\DeclareSIUnit\angstrom{\text{Å}}

\title{Missing Wedge Inpainting and Joint Alignment in Electron Tomography through Implicit Neural Representations}

\author[1, *, $\dagger$]{Cedric Lim}
\author[2, *, $\ddag$]{Corneel Casert}
\author[1]{Arthur R. C. McCray}
\author[1]{Serin Lee}
\author[3]{Andrew Barnum}
\author[1]{Jennifer Dionne}
\author[1, $\star$]{Colin Ophus}

\affil[1]{Department of Materials Science and Engineering, Stanford University, 496 Lomita Mall, Stanford, 94305, CA, USA}
\affil[2]{National Energy Research Scientific Computing Center, Lawrence Berkeley National Laboratory, 1 Cyclotron Road, Berkeley, 94720, CA, USA}
\affil[3]{nano@Stanford Shared Facilities, Stanford University, 348 Via Pueblo, Stanford, 94305, CA, USA}

\affil[$\dagger$]{cedlim@stanford.edu}
\affil[$\ddag$]{ccasert@lbl.gov}
\affil[$\star$]{cophus@stanford.edu}
\affil[*]{These authors contributed equally to this work}

\begin{abstract}
    Electron tomography is a powerful tool for understanding the morphology of materials in three dimensions, but conventional reconstruction algorithms typically suffer from missing-wedge artifacts and data misalignment imposed by experimental constraints.
    Recently proposed supervised machine-learning-enabled reconstruction methods to address these challenges rely on training data and are therefore difficult to generalize across materials systems.
    We propose a fully self-supervised implicit neural representation (INR) approach using a neural network as a regularizer.
    Our approach enables fast inline alignment through pose optimization, missing wedge inpainting, and denoising of low dose datasets via model regularization using only a single dataset.
    We apply our method to simulated and experimental data and show that it produces high-quality tomograms from diverse and information-limited datasets.
    Our results show that INR-based self-supervised reconstructions offer high fidelity reconstructions with minimal user input and preprocessing, and can be readily applied to a wide variety of materials samples and experimental parameters.
\end{abstract}
\begin{document}
\maketitle

\section{Introduction}\label{sec1}

Electron tomography (ET) has emerged as a powerful method for resolving the three-dimensional (3D) structure of materials and biological specimens at the nanoscale. 
In materials science, tomographic quantification of nanoparticle size distribution, morphology, and composition has been used to explain electrocatalytic properties \cite{wang_recent_2022, xia_bimetallic_2018, lee_bimetallic_catalyst, mckeown_green_catalyst}, while atomic electron tomography (AET) enables determination of 3D atomic positions and species for direct evaluation of catalytic activity \cite{jeong_atomic-scale_2025, xiang_3d_2022, pelz_simultaneous_2022, lee_single-atom_2021}.
In structural biology, cryo-electron tomography (cryo-ET) has been instrumental in resolving the 3D architectures of viruses, ribosomes, and complex biomolecular assemblies \cite{cressey_cryo-electron_2017, turk_promise_2020, nogales_development_2016, oikonomou_new_2016}.

Tomography datasets are collected by taking a series of images while tilting the sample through a range of discrete tilt angles~\cite{ercius_electron_2015, ophus_3d_2019}.
Many standard reconstruction algorithms were originally developed for X-ray computed tomography \cite{schulz_how_2021}, where the specimen is rotated between the source and detector and projections are typically acquired over a full $180^\circ$ (parallel-beam) or $360^\circ$ (cone-beam) rotation \cite{kak2001principles}.
For most materials samples, dense angular sampling is often feasible because X-rays cause relatively little radiation damage during acquisition \cite{du_plessis_comparison_2016}.
In contrast, for datasets collected using (scanning) transmission electron microscopy ((S)TEM), the accessible tilt range (typically $\sim\pm70^\circ$) is restricted by the geometry of the microscope pole piece.
The incomplete angular coverage leads to a missing-wedge of information in reciprocal space that greatly degrades reconstructions.
Moreover, coarse angular sampling is often necessary in electron tomography to minimize beam damage, resulting in small missing-wedge artifacts between projections \cite{leijten2017quantitative}.
Reconstruction fidelity is further reduced by alignment-related errors, including uncertainty in the tilt axis and relative shifts between projections \cite{kubel2005recent}.
By comparison, X-ray tomography typically uses mechanically stable rotation stages and is less affected by alignment issues \cite{sun2016reference}, or fiducial markers can be easily incorporated to refine alignment \cite{parkinson2012automatic}.

The most widely used conventional reconstruction algorithms, such as filtered back-projection (FBP) \cite{frank_three-dimensional_2006, momose_phase-contrast_1996} and the simultaneous iterative/algebraic reconstruction techniques (SIRT/SART) \cite{gilbert_iterative_1972, andersen_simultaneous_1984}, have strict data requirements to achieve high-quality reconstructions. 
FBP requires a large set of high-quality, low-noise projections for accurate results, which is generally feasible in X-ray tomography. 
Iterative methods such as SIRT, while more computationally demanding, are more tolerant of limited or noisy data and therefore remain the primary choice for most electron tomography experiments. 
Inline alignment is often used to correct translational shifts, but it is generally limited to coarse refinements.
More recently, algorithms developed specifically for electron tomography have addressed tilt-axis misalignment and missing-wedge artifacts using Fourier-based approaches \cite{miao_equally_2005, pham_accurate_2023, noauthor_genfire_nodate}.
However, these methods are often constrained by the computational cost of repeated Fourier and inverse Fourier transforms and by slow convergence.

In recent years, deep learning has become the focal point in improving tomographic reconstructions, particularly for interpolating the missing wedge in tomographic datasets.
In medical X-ray tomography, numerous methods based on implicit neural representations (INR) have been developed for denoising and sparse-view reconstruction \cite{gao_h-siren_2024, zhu_finer_2024, mildenhall_nerf_2020, wang_sparsenerf_2023, zha_r2-gaussian_2024, cai_radiative_2024}; for instance, representing the volume using a neural-network model effectively reduces streaking artifacts in computed tomography (CT) \cite{wang_neural_2024, veen_missing_2024}. 
In cryo-electron tomography (cryo-ET), diverse approaches have emerged, including CNN-based subtomogram inpainting \cite{liu_isotropic_2021}, iterative angular masking \cite{wiedemann_deep_2024}, and frame interpolation between tilt images \cite{majtner_cryotiger_2025}.
These methods typically rely on subtomogram averaging and supervised learning under assumptions of structural uniformity.
Overall, most deep learning–based reconstruction methods remain dependent on fully supervised or semi-supervised training, often requiring tomographic data from similar nanoparticles for network pre-training \cite{yao_no_2024}.
Consequently, the reliance on supervised learning renders most deep learning–based reconstruction techniques unsuitable for materials science applications, where samples are heterogeneous and collecting multiple tomograms is impractical.
In addition, the majority of these machine learning algorithms do not optimize for tilt-axis or translational misalignments, imposing an additional time-consuming requirement of high-quality manual alignment.

In this work, we develop a fully self-supervised and easily parallelizable tomographic reconstruction algorithm that jointly performs alignment and missing-wedge inpainting using INRs.
We use a neural-network model to represent the volume as a continuous physical shape function.
This framework enables denoising of heavily dose-limited datasets but also recovers missing-wedge information without relying on any explicit sinogram inpainting. 
In addition, the framework robustly reconstructs sparsely sampled data, effectively compensating for small missing wedges and producing high-fidelity reconstructions of dose-sensitive samples. 
By jointly learning the full 3D pose of the object corresponding to each tilt image, our method achieves accurate reconstructions without the need for prior fine alignment. 
We apply INR tomography with joint alignment to experimental tilts series from two different materials systems, and a simulated phantom to directly compare with conventional algorithms showing marked improvements. 
We have implemented this algorithm in the open-source Python package \verb|quantEM|.
INR tomography provides high-quality reconstructions and automated alignment for a wide range of materials tomography datasets.

\section{Results}\label{sec11}

\begin{figure}[h!]
    \centering
    \includegraphics[width=1\linewidth]{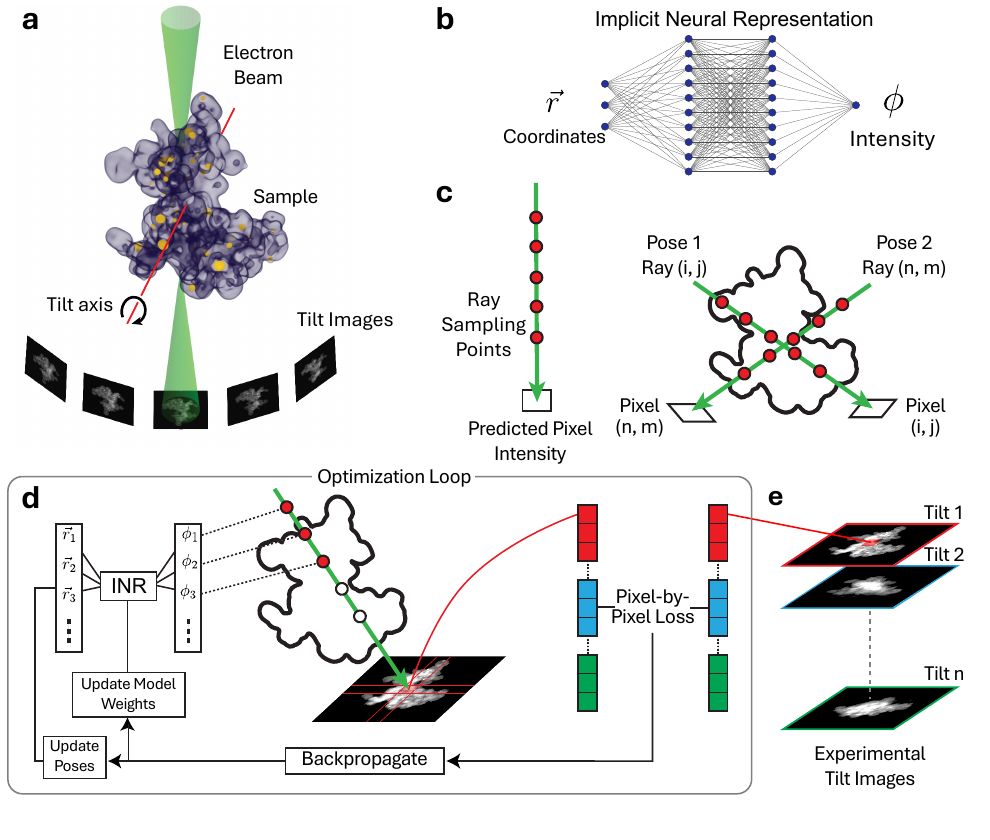}
    \caption{\textbf{Overview of the INR-based tomographic reconstruction algorithm.} \textbf{(a)} Schematic of a tomographic experiment using STEM. An image is recorded at each tilt step as the sample is rotated around the tilt axis. 
    \textbf{(b)} A neural network receives a 3-D position vector, $\vec{r} = \{x, y, z\}$ as input, and produces an intensity value, $\phi$, at that position. This way of representing the volume is called an implicit neural representation (INR).
    \textbf{(c)} The reconstruction volume is sampled by rays that terminate at individual pixels of the experimental data. 
    Each ray also incorporates and optimizes the pose for each tilt image in the dataset.
    \textbf{(d)} Optimization loop schematic. Coordinates sampled along each ray are fed to the INR, whose outputs are integrated to predict the pixel intensity. This prediction is compared to the experimental measurement with a pixel-wise loss, which is back-propagated to update both the INR weights and the pose.
    \textbf{(e)} Schematic showing how the experimental tilt images are reshaped into a 1D array of pixels that are compared to individual pixel predictions made using the INR.}
    \label{fig:main_figure}
\end{figure}

\subsection{INR Tomography Workflow}\label{sec:workflow}

A schematic of our reconstruction workflow is shown in Fig. \ref{fig:main_figure}. 
The sample is rotated about an arbitrary tilt axis with a specified tilt step, where an image is collected per angle. 
During acquisition, the sample is roughly re-centered and rotation is nominally along one axis, but in practice the holder deviates from the ideal tilt axis due to inherent runout resulting from manufacturing tolerances and imperfections. (Fig. \ref{fig:main_figure}a).
Unlike conventional algorithms that use only a single tilt angle for each image and estimate the image shifts as a pre-processing step, our method solves for the complete 3D orientation and in-plane translation, i.e.~the full sample pose of each tilt image.

Because conventional algorithms generally do not optimize orientation parameters, a natural machine-learning approach is to represent the object on a voxelized 3D grid, project it for each pose, and define a loss from the difference between predicted and measured image intensities.
The gradients of both the imaging parameters and object representation can be solved via automatic differentiation (AD). 
In this method, one would be able to perform gradient descent and find the ground truth tilt axis and relative shifts for each projection. 

However, this approach requires a fully differentiable projection operator that rotates the object to arbitrary orientations, and each rotation involves interpolating the voxelized volume.
Most importantly, parallelization of the projection operator is difficult as the volume would have to be instantiated across multiple GPUs in order to rotate it to a specific pose.

Rather than using a voxel grid to represent the volume, we instead use an INR as a model regularizer to represent the volume.
Specifically, we use a fully-connected neural network with periodic activation functions \cite{gao_h-siren_2024}, which takes as input an arbitrary coordinate vector and returns an intensity value (Fig. \ref{fig:main_figure}b). 
In our testing, this architecture had the least amount of hyperparameter tuning needed to converge to high spatial frequencies, in contrast to other architectures \cite{zhu_finer_2024}.
Additionally, using an INR allows for arbitrarily fine sampling of the volume, upsampling, and easy modification of the support constraint \cite{scott_electron_2012}.
Most importantly, we are able to project the volume at any orientation without interpolation by using precisely rotated projection vectors corresponding to each pose. 
We first compute rays along the beam direction with discrete sampling points (Fig. \ref{fig:main_figure}c) which correspond to each pose. 
By integrating the intensity along a ray, we obtain a predicted pixel intensity that we can compare directly with the experimentally collected tilt projection, pixel-by-pixel. 
This simple ``forward model'' of integration works for HAADF STEM, as to a first approximation images collected in this modality depend linearly on the projected potential of the sample \cite{ophus_3d_2019}. 
With the predicted intensity value, we can compute a pixel-by-pixel smooth$_{L_1}$ loss (see Methods), which is back-propagated to simultaneously optimize the model weights and poses until the loss converges (Fig. \ref{fig:main_figure}d). 
To speed up convergence, one can perform a rough initial alignment and pre-train the network on a fully converged SIRT reconstruction.
To show that this method is truly self-supervised, we initialize the network randomly without any pre-training for the rest of this work. 
Each pixel can be evaluated independently, allowing for straightforward parallelization over many GPUs.
The training procedure is described in detail in Sec.~\ref{sec:methods-optimization}.

\subsection{Experiments with a simulated phantom}\label{sec:phantom}

\begin{figure}[t!]
    \centering
    \includegraphics[width=.865\linewidth]{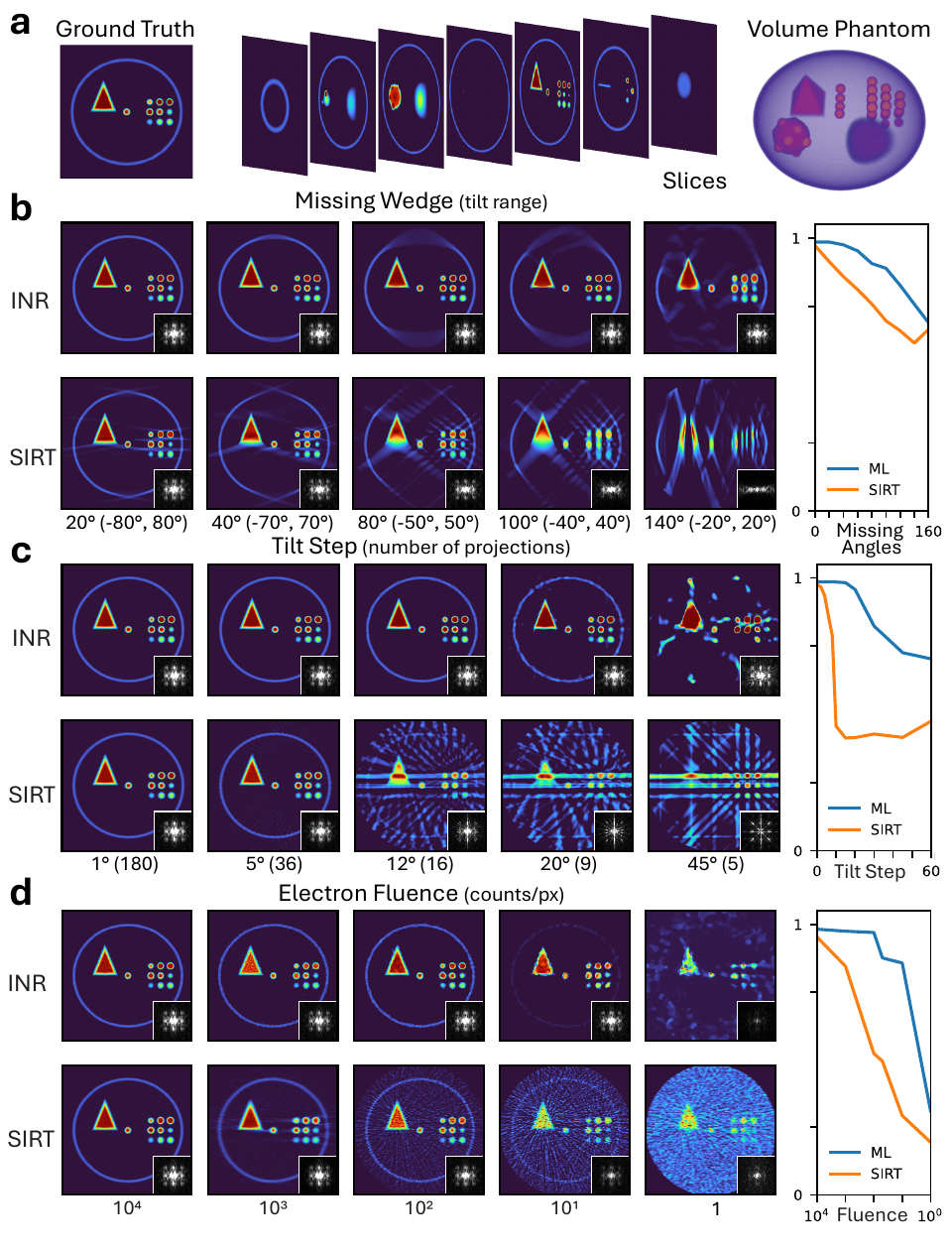}
    \caption{\textbf{Comparison between INR and SIRT for a simulated phantom}. \textbf{(a)} Reference slice of the ground truth phantom, slices along the volume, and a 3D render. \textbf{(b-d)} Comparison between our method (INR) versus SIRT by varying the missing wedge size, tilt step, and electron fluence respectively through a slice perpendicular to the missing wedge direction. For each case we also compute the SSIM between each reconstructed slice to the ground truth.
    }
    \label{fig:phantom_results}
\end{figure}

By representing the volume with an INR, the neural network acts as an implicit prior that constrains the volume, leading to a more physical reconstruction even for challenging datasets (e.g noisy, sparse, or misaligned). 
We first demonstrate this for a simulated 3D phantom, shown in Fig.~\ref{fig:phantom_results}a, and compare the INR reconstruction to a conventional iterative reconstruction algorithm (SIRT).
The phantom contains internal 3D features with varying intensities and structures with shapes including triangular and spherical volumes, all wrapped in a thin ellipsoid membrane. 
We perform a series of experiments to demonstrate the effectiveness of our method: varying the missing wedge size, changing the tilt step between projections, and reducing the electron fluence. 
For each test case, we plot an image which averages a 5-voxel thick slice through the center of the volume, with the missing wedge direction shown vertically.
We have inset the Fourier transform magnitude of each slice in the lower right corner.
We evaluate the reconstruction quality by using the structural similarity index measure (SSIM) \cite{wang_image_2004} comparing each reconstructed volume to the ground truth volume shown in Fig.~\ref{fig:phantom_results}a. 

We first vary the missing-wedge extent, which is shown in Fig.~\ref{fig:phantom_results}b. 
We used a $1^\circ$ tilt increment for all panels.
The leftmost slices shown in Fig.~\ref{fig:phantom_results}b were reconstructed from data using a tilt range of $(-80^\circ,80^\circ)$, i.e. a $20^\circ$ missing wedge, and have SSIM values equal to 0.985 and 0.972 for the INR and SIRT reconstructions respectively. 
For a small missing wedge, the INR is able to regularize the reconstruction, while SIRT begins to show streaking artifacts between features, and both elongation along with missing segments of the membrane.
As we increase the missing wedge angle, the SIRT reconstructions degrade significantly, and the internal structure is dramatically different with an $80^\circ$ missing wedge.
The INR in contrast is able to reconstruct the internal features accurately up to a missing wedge of $100^\circ$, and degrades gracefully at even larger missing wedge angles.
The INR reconstructs the membrane more accurately than SIRT across all missing wedge angles, although gaps remain along the missing wedge direction at larger angles.
The benefit of using an INR reconstruction is also clearly shown in the inset Fourier transform magnitudes, where the large missing wedge is clearly visible in the SIRT reconstructions.
The INR is able to learn approximate shape functions which effectively inpaint the missing wedge without any prior training.

Next we vary the tilt step while maintaining the full tilt range of $(-90^\circ,90^\circ)$, as shown in Fig.~\ref{fig:phantom_results}c.
The INR is much more robust and can generate accurate volumes of up to $20^\circ$ step size, while SIRT significantly suffers with tilt steps of greater than $5^\circ$.
Going to larger tilt steps, the missing information between each tilt become increasingly apparent in the Fourier transform magnitudes.
In particular, the INR recovers the ground-truth volume using only 9 projections, provided high-angle tilts are included, with SSIM values of 0.957 for INR and 0.427 for SIRT.
The SSIM scores for SIRT drop sharply once the tilt step exceeds $10^\circ$, whereas the ML reconstruction maintains an average SSIM of 0.981 up to 20°, compared with 0.683 for SIRT.
Since this experiment has no missing wedge, these results highlight that extending the tilt range to higher angles improves reconstruction quality more than simply reducing the tilt step size.

In TEM imaging, the signal-to-noise ratio is often limited by electron fluence; when operating at low dose, Poisson noise dominates the reconstruction.
We vary electron fluence by adding Poisson noise to each tilt image, keeping a 1° tilt step and no missing wedge, as shown in Fig.~\ref{fig:phantom_results}d.
The INR is capable of denoising the projections and is able to reconstruct a much smoother volume compared to SIRT. 
At a fluence of $10^4$ counts/px, SIRT and INR are practically indistinguishable, however the reconstruction quality degrades rapidly when the fluence is decreased.
The SIRT reconstruction is visibly noisy for a fluence of 100 counts/px, while the INR reconstruction is almost noise-free.
At a fluence of 1 count/px, the INR still resolves the internal structures, while SIRT fails to reconstruct a meaningful volume.
However, the INR does begin to overfit to noise, and so we perform holdout validation as a benchmark for early-stopping; in this case outputting the reconstructed volume when the loss is minimal on the holdout set (see Sec.~\ref{sec:methods-optimization}).
This trend is reflected in the SSIM, which drops sharply with dose; at a dose of 10 count/px, the SSIM is 0.857 for INR and 0.293 for SIRT. 

\begin{figure}[h!]
    \centering
    \includegraphics[width=1\linewidth]{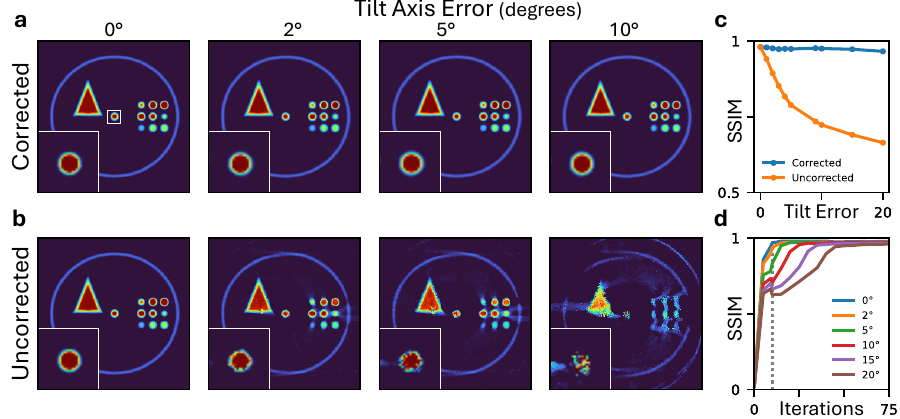}
    \caption{\textbf{INR corrected and uncorrected pose comparison.} \textbf{(a)} Slice through the phantom volume for varying tilt-axis error pose optimization, \textbf{(b)} without pose optimization. The inset shows the zoomed in circular feature in the middle of the slice. \textbf{(c)} SSIM for different tilt-axis errors for the corrected and uncorrected case. \textbf{(d)} SSIM and the number of training epochs required to converge depending on the degree of the tilt-axis error. The gray dashed line indicates where we begin optimization of the poses.}
    \label{fig:phantom_alignment}
\end{figure}

To benchmark our inline alignment routine, we deliberately misalign the tilt stack by applying a systematic tilt axis error in Fig.~\ref{fig:phantom_alignment}.
As discussed in Sec. \ref{sec:workflow}, we directly optimize the full 3D orientation of the volume corresponding to each tilt image.
In Fig.~\ref{fig:phantom_alignment}a, it is clear that our pose optimization is robust to high degrees of rotational misalignment.
Without incorporating any inline alignment in Fig.~\ref{fig:phantom_alignment}b, the reconstruction degrades considerably even at small tilt-axis values.
Similar to conventional algorithms, the volume begins to exhibit streaking artifacts along with high frequency artifacts.
This is highlighted in the SSIM in Fig. ~\ref{fig:phantom_alignment}c decreasing significantly with increasing amounts of misalignment.
In contrast, by optimizing the pose we are able to retrieve the ground truth structure no matter the degree of rotational misalignment achieving an average SSIM of .974.
However we note that large degrees of misalignment correspond to longer convergence times as seen in Fig.~\ref{fig:phantom_alignment}d.
While it is feasible to let the INR perform the full alignment, it is still recommended to perform a coarse preprocessing routine to speed up convergence time.

\subsection{Experimental tomography of catalytic nanoparticles}\label{sec:catalytic_nanoparticle}

To experimentally validate and test our methods, we have used INR and SIRT to reconstruct tilt series recorded using a STEM instrument.
We used the high annular dark field (HAADF) imaging mode, as it produces approximately linear projected contrast \cite{ophus_3d_2019}.
The first sample we reconstructed consists of AuPd nanoparticles on a TiO$_2$ support \cite{lee_bimetallic_catalyst}.
The nanoparticles are the minority phase, dispersed over the surface of the support.
We collected projections by interleaving two tilt series with $4^\circ$ steps to make one singular tilt series with a range of $\pm70^\circ$ and a tilt step of $2^\circ$. 
Prior to reconstruction we applied a polynomial fit to  each projection for background subtraction, fitting to the areas outside of the sample (see Sec.~\ref{sec:method_catalytic})

\begin{figure}[htb]
    \centering
    \includegraphics[width=1\linewidth]{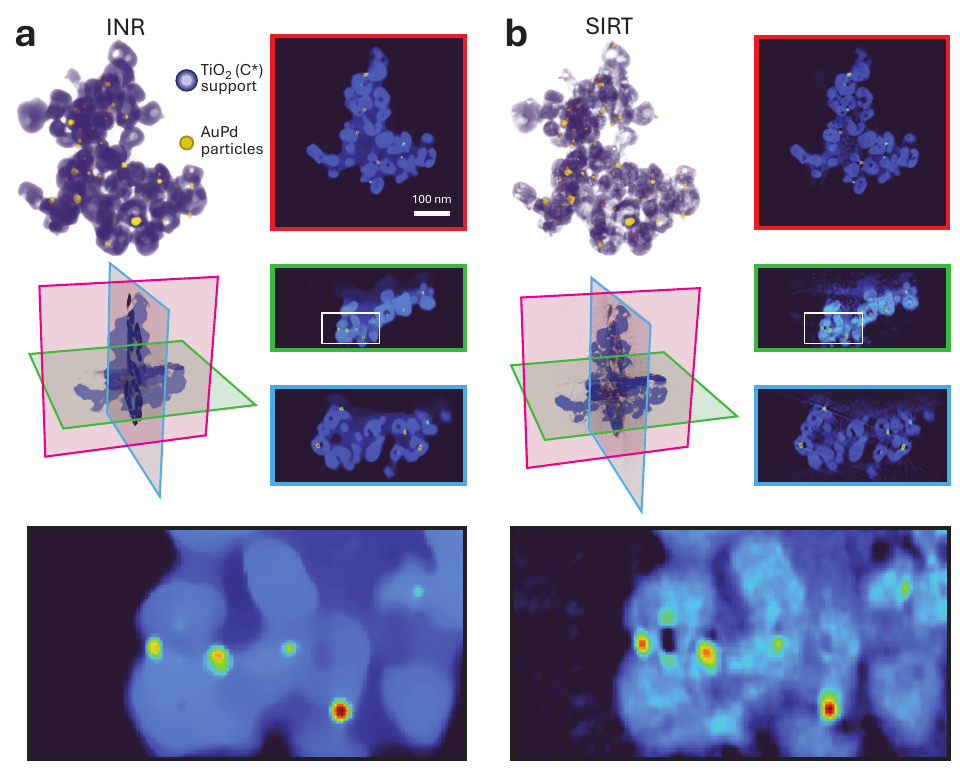}
    \caption{\textbf{INR and SIRT tomographic reconstructions of catalytic AuPd nanoparticles on a TiO$_2$ support}. A 3D visualization and image slices of the reconstructed volume using \textbf{(a)} INR and \textbf{(b)} SIRT. The right panels show X-Y, X-Z, and Y-Z slices from top to bottom respectively. The bimodality seen is suspected to be carbon contamination (see Supp. Info).}
    \label{fig:experimental_sirtvsml}
\end{figure}

As discussed in Sec.~\ref{sec:phantom}, by jointly optimizing the full 3D orientation of each projection the model learns self-consistent alignment parameters across the entire series.
Applying this approach to the experimental dataset yields a substantially higher-fidelity reconstruction than SIRT, as shown in Fig. \ref{fig:experimental_sirtvsml}a.
The SIRT reconstruction (Fig. \ref{fig:experimental_sirtvsml}b) exhibits visible missing wedge and misalignment artifacts, seen as streaking and elongation of individual nanoparticles, whereas the INR isolates each particle and refines the surrounding support with well-defined boundaries.
Although the SIRT reconstruction appears reasonable, the INR reveals sharper features throughout the volume, including a clearly resolved lighter contrast region from the carbon support and possible carbon contamination in some sample regions.
We have confirmed the presence of carbon using x-ray energy-dispersive spectroscopy (XEDS) measurements shown and discussed in Supplemental Figure \ref{fig:catalytic_eds}.

\begin{figure}[htb]
    \centering
    \includegraphics[width=1\linewidth]{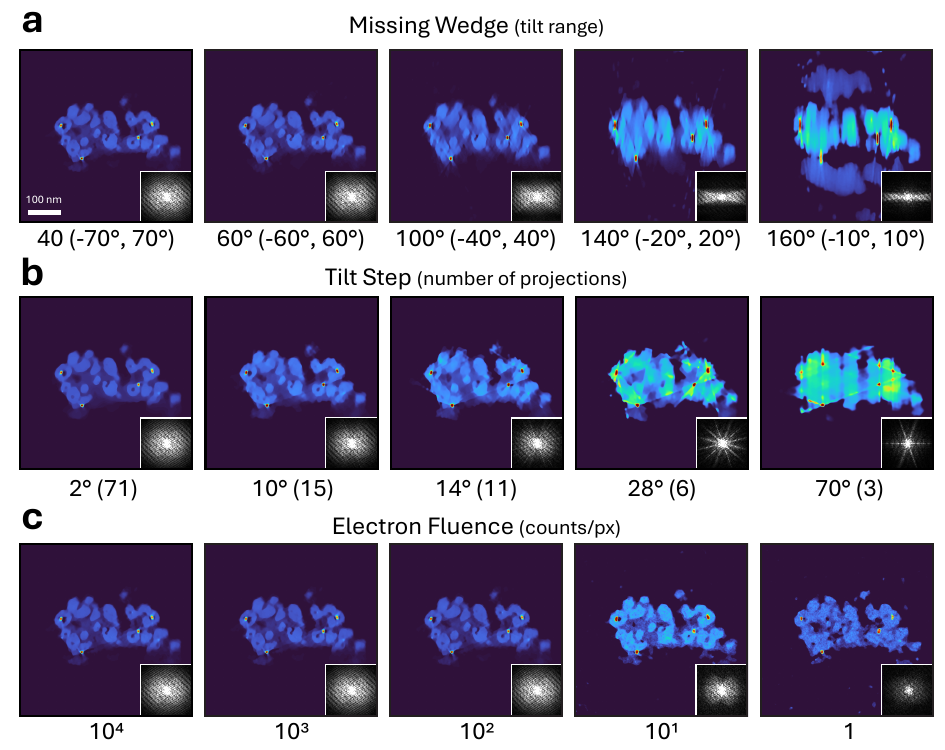}
    \caption{\textbf{Supported nanoparticles INR reconstructions for different experimental parameters}.
    \textbf{(a)–(c)} Reconstructed slices perpendicular to the missing wedge direction for varying missing wedge sizes, tilt steps, and electron fluences, respectively.
    }
    \label{fig:experimental_results}
\end{figure}

We next test the robustness of our method on experimental data to evaluate how the trends observed in the simulated phantom extend to real measurements.
We perform the same set of experiments as in Sec.~\ref{sec:phantom}, varying the missing wedge while keeping a $2^\circ$ tilt step and examining slices along the missing wedge direction as shown in Figure~\ref{fig:experimental_results}a.
Compared to the simulated phantom, this experimental dataset is more challenging since the INR must also infer information lost within the small gaps between projections caused by the $2^\circ$ tilt step.
Consistent with the simulated results, the INR produces accurate reconstructions up to a missing wedge of $100^\circ$ ($\pm 40^\circ$ tilt range).
At more limited tilt ranges, the nanoparticles remain identifiable but exhibit increasing vertical distortion and missing wedge artifacts.

To further evaluate our method on sparsely collected datasets, we now vary the tilt step while maintaining the original tilt range of the dataset as shown in Fig.~\ref{fig:experimental_results}b.
Since we originally start with a limited tilt range, we use values that provide symmetric angular coverage in the positive and negative tilt angle ranges.
Tilt steps between $2^\circ$ and $10^\circ$ were not included since there was no considerable changes to the volume.
Similar to the phantom, this test emphasizes the importance of prioritizing higher angular coverage over tilt steps.
Even with $10^\circ$ steps our algorithm is still able to resolve the general structure and particles within the volume.
However, subsequent increases to the tilt step produces the same mini-missing wedge artifact seen with SIRT.
Surprisingly, even with only 11 projections we can still clearly resolve the nanoparticles and with only 6 projections we could estimate the approximate position of the 4 nanoparticles visible in this slice. 

Since we cannot experimentally change the electron fluence post-acquisition, we instead simulate it by adding Poisson noise similar to Sec.~\ref{sec:phantom}.
As observed in the phantom, the INR is incredibly resilient to low-dose conditions with no appreciable change until a dose of $10$ counts/pixel.
At lower doses, the INR begins to fit to noise and performing holdout validation becomes necessary in these conditions. 

\subsection{Experimental tomography of a hyperbranched nanoparticle}

\begin{figure}[htb]
    \centering
    \includegraphics[width=1\linewidth]{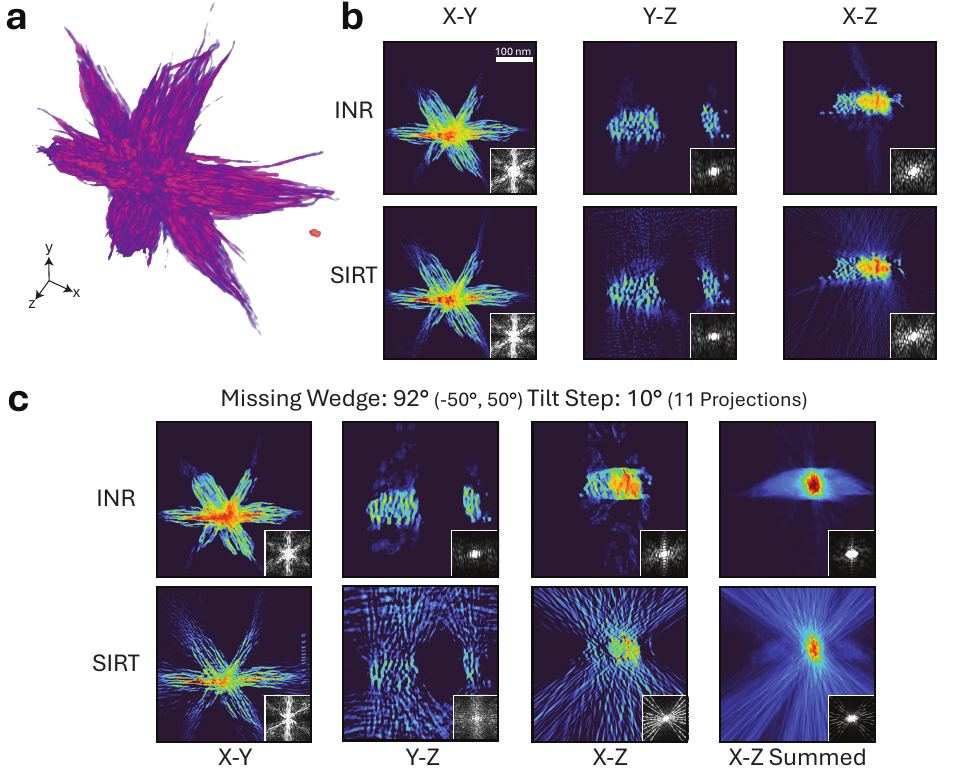}
    \caption{\textbf{INR and SIRT reconstructions of a hyperbranched Co$_2$P nanoparticle}. \textbf{(a)} 3D rendering of the sample using an INR. \textbf{(b)} Slices from the INR and SIRT at X-Y, Y-Z, and X-Z projections. \textbf{(c)} Slices from SIRT and INR reconstructions across different projection planes, with a simulated $92^\circ$ missing wedge and $10^\circ$ tilt steps.}
    \label{fig:ml_nanostar_sirt_vs_ml}
\end{figure}

We reconstruct a hyperbranched Co$_2$P nanoparticle with approximate six-fold symmetry and an included fiducial to study the effect of high spatial frequency and low intensity features in the tilt series published by Levin et al. \cite{levin_nanomaterial_2016} (Fig. \ref{fig:ml_nanostar_sirt_vs_ml}).
As discussed in \cite{levin_nanomaterial_2016}, the tilt range of this sample is from $\pm75^\circ$ with a $2^\circ$ step, with the $23^\circ$ and $25^\circ$ angles removed due to a goniometer backlash. 
The previous results feature relatively simple structures, allowing the INR to recover physical shape functions with ease.

In Fig. \ref{fig:ml_nanostar_sirt_vs_ml}a, we show a 3D render of the sample using our algorithm.
We also provide a direct comparison between the INR and SIRT reconstruction methods in Fig. \ref{fig:ml_nanostar_sirt_vs_ml}b, where for both methods we used the preprocessed dataset provided in the original dataset and performed background subtraction. 
In Fig.~\ref{fig:ml_nanostar_sirt_vs_ml}b, the additional pose refinements and INR regularization reveal finer high-frequency details and suppress the streaking artifacts that are prominent in the SIRT reconstruction.
Moreover, it is clear that the INR is not heavily affected by missing wedge and small missing wedge artifacts.
In the FFT inset of the SIRT X-Z slice, the two missing projections from the tilt series are visible, whereas they are not apparent in the INR reconstruction.

For this sample we simulate limited experimental conditions by restricting the tilt series to a maximum angle of $\pm 55^\circ$ with a tilt step of $10^\circ$ (Fig. \ref{fig:ml_nanostar_sirt_vs_ml}c).
Effectively, this reduces the total dose by a factor of 7 throughout the whole tilt series.
In practice, this could happen if one were to collect an interleaved tilt series with a step of $5^\circ$ and the sample becomes too beam damaged to perform the last tilt series. 
With SIRT, the reconstruction quality is clearly limited by the small number of available projections.
Although the X–Y slices still yield a reasonable reconstruction, the resolution along the other axes is strongly degraded by missing information and algorithm-induced streaking artifacts.
In the X-Z summed projection, the structure is heavily deteriorated and not interpretable. 
However, the INR reconstruction is able to reduce the artifacts observed in SIRT and is still is able to retrieve a reasonable tomogram. 
Most importantly, even with 11 projections the INR is still robust to both types of missing wedge artifacts.
However, comparing to the best case scenario with all projections, from a limited number of projections the INR shows significant smearing in the vertical directions of the Y-Z and X-Z slices.
Despite this, the INR can produce reasonable quality reconstructions even with heavily constrained experimental parameters that would be difficult to achieve methods that do not incorporate machine learning.

\begin{figure}[htb]
    \centering
    \includegraphics[width=1.0\linewidth]{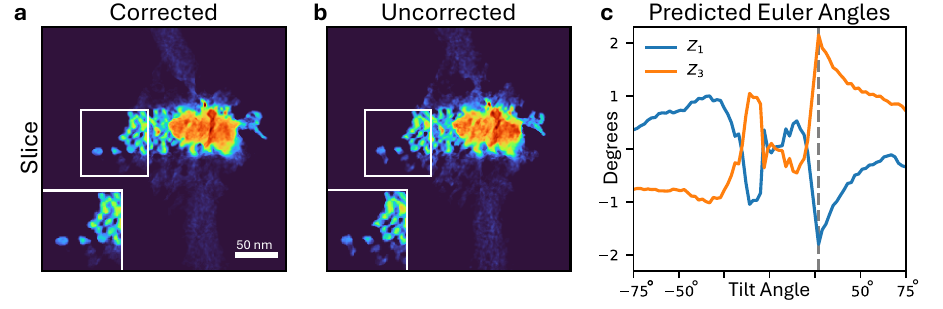}
    \caption{\textbf{INR reconstructions of a hyperbranched Co$_2$P nanoparticle with alignments corrected and not corrected}. Slice of the INR reconstruction \textbf{(a)} with pose optimization \textbf{(b)} and without optimization. \textbf{(c)} The predicted $Z_1$ and $Z_3$ Euler angles. The gray line corresponds to the goniometer backlash observed in \cite{levin_nanomaterial_2016}.}
    \label{fig:nanostar_alignment}
\end{figure}

For this sample, alignment is a challenge because fine corrections are required for each tilt projection. 
In Fig.~\ref{fig:nanostar_alignment}, we compare INR reconstructions with (corrected) and without (uncorrected) pose optimization, using the same reconstruction parameters in both cases.
Comparing both Fig.~\ref{fig:nanostar_alignment}a and Fig.~\ref{fig:nanostar_alignment}b, we are able to minimize the misalignment induced smearing artifact and better isolate individual branches.
Additionally, our algorithm is able to correct for the goniometer backlash in the experiment as seen in Fig.~\ref{fig:nanostar_alignment}c.
Together, these improvements demonstrate that our pose-optimization methodology can reliably recover high-fidelity structural information and is able to correct for experimental errors.

\section{Discussion}

We have developed a self-supervised and parallelizable tomography reconstruction algorithm that simultaneously aligns, denoises, and fills in some of the lost information due to the missing wedge artifact.
In comparison to current iterative and machine learning tomography methods, using INRs for tomographic reconstructions provides improved robustness to information-limited datasets, while improving reconstruction quality for experimental datasets.
Additionally, the INR does not require supervised training for missing wedge inpainting, and can allow better tomographic experiments of beam-sensitive materials through denoising. 
Most importantly, our algorithm automates projection alignment, which is often the most difficult and time-consuming part of tomography. 
Allowing for learning the full 3D pose of each projection is key to obtaining accurate tomograms. 
Furthermore, we are able to reconstruct large-scale volumes due to the parallelizable nature of INRs.

Reconstruction speed is highly dependent on the number of GPUs available in this framework.
The experiments conducted in this work were performed using a total of 16 GPUs per reconstruction. For both experimental cases using the full dataset for 150 epochs took 2 hours.
Without access to multi-GPU systems, and depending on the volume size, reconstructions speed will take considerably longer.
Nonetheless, since INR iteration speed scales nearly linearly with the number of GPUs this problem can be effectively mitigated.
This work also highlights that reconstruction quality is highly sensitive to the choice of loss function (See Sec.~\ref{sec:methods-optimization} and Supplemental Information Sec.~\ref{supp_sec:l1l2comp}). 
Nevertheless, we have proposed guidelines and how to use cross-validation to reconstruct a variety of samples which may minimize the need for performing a hyperparameter search.

We anticipate that multimodal and low-dose experiments, such as XEDS tomography, will benefit substantially from our INR framework, as radiation damage from repeated chemical mapping often limits achievable resolution and reconstruction quality.
Furthermore, conventional paralell-beam TEM imaging contains phase and diffraction contrast that convey useful structural information.
Currently, most existing algorithms do not incorporate this into the physics forward model, leading to degraded reconstruction quality. 
Our framework may enable the direct inclusion of this underlying imaging physics, overcoming these limitations.
Additionally for atomic-resolution electron tomography (AET), where precise alignment and faithful reconstruction of atomic potentials are critical, our approach may offer several advantages over existing methods.
Together, this reconstruction framework based on inline alignment and INRs could significantly enhance electron tomography as a whole, enabling higher-quality reconstructions with fewer projections and less stringent experimental requirements.

%TC:ignore
\section{Methods}

\subsection{Reconstruction implementation}\label{sec:methods-optimization}

\subsubsection{Volume representation}

The volume is represented through an implicit neural network, which maps coordinates $\mathbf{r} = \{x,y,z\} \in [-1,1]^3$  to a value for the intensity $\phi$. Such a continuous representation has several advantages over grid-based methods, including being more memory-efficient for large volumes (the number of neural-network weights can be much smaller than the number of voxels in a discrete representation), allowing for analytical calculation of derivatives, and not requiring interpolation to evaluate a shifted or rotated volume. 

We use an INR with periodic activation functions, which are less affected by the spectral-bias problem (i.e. that low-frequency components are more easily learned) than ReLU-based neural networks \cite{gao_h-siren_2024}. 
Specifically, we use the H-Siren network, which is a fully-connected network with as activation function for the input layer \begin{equation}
    \sigma_{\textrm i}(x) = \sin(\omega_0 \sinh(2x)),
\end{equation} and for the hidden layers
\begin{equation}
    \sigma_{\textrm h}(x) = \sin(\omega_0 x),
\end{equation}
where $\omega_0$ is a hyperparameter that controls the frequency and is kept fixed at $\omega_0 = 30$. 
To obtain a nonnegative value for the predicted intensity $\phi$, we use a softplus output activation function
\begin{equation}
    \sigma_{\textrm{o}}(x) = \ln(1+\exp(x)). 
\end{equation}
Throughout this paper, we use an H-Siren network with four hidden layers and a hidden dimension of 512. We initialize the weights $\{w\}$ for the input layer as \begin{equation}
    w^{\rm i} \sim U(-1/n, 1/n),
\end{equation}
where $n$ is the number of input features to the layer (``fan-in''), and for the other layers as 
\begin{equation}
    w^{\rm h} \sim U\left(-\sqrt{6/n}/\omega_0, \sqrt{6/n}/\omega_0\right).
\end{equation}

\subsubsection{Optimization}\label{sec:optim}
To optimize the volume representation and poses (shifts and tilts), we minimize the loss function
\begin{equation}
    \mathcal{L} = \mathcal{L}_{\textrm{pixel}} + \alpha \mathcal{L}_{\textrm{TV}},\label{eq:loss}
\end{equation} 
where the first term is the $\text{smooth}_{\mathcal{L}_1}$ loss \cite{gao_h-siren_2024, girshick_fast_2015} of the predicted and experimentally measured pixel intensities. The second term is a regularization term, and $\alpha$ is a hyperparameter which determines the weight of the regularization term which is a tunable parameter that needs to be adjusted depending on the sample.

The first term is given by
\begin{equation}
    \mathcal{L}_{\textrm{pixel}} = \frac{1}{M N^2} \sum_{j=1}^M\sum_{i=1}^{N^2}\textrm{smooth}_{L_1}(I^{i,j}_{exp} - I^{i, j}_{pred})
\end{equation}
where $I_{\textrm{exp}}^{i,j}$ is the intensity of pixel $i$ in the $N\times N$ tilt series image $j$, of which there are $M$. The $\textrm{smooth}_{L_1}$ is defined as,
\begin{equation}
    \textrm{smooth}_{L_1}(x) = \begin{cases}
        \frac{0.5x^2}{\beta} & \textrm{if }|x|<1 \\
        |x|-0.5\beta & \textrm{otherwise}
    \end{cases}
\end{equation}
As described in \cite{girshick_fast_2015}, we use a value for the hyperparameter $\beta=0.07$ which defines the crossover point between using $\mathcal{L}_1$ or $\mathcal{L}_2$ loss. This choice of $\beta$ was the most robust for showing low and high spatial frequencies, and low intensity features across all experiments (see Supplemental Information Sec.~\ref{supp_sec:l1l2comp}). This parameter may also be tuned to enhance reconstructions. 

To calculate the predicted pixel intensity $I_{\textrm{pred}}^{i,j}$, we need to integrate the intensity $\phi$ along a line (``ray'') which ends at pixel $i$. This ray can be found by rotating and shifting a ray going along the $z$ axis in the original coordinate system according to the tilt angle $\theta^j$ and the pose 
(tilt axis oscillation and shifted position), see Fig.~\ref{fig:main_figure}c. We parameterize the latter for each tilt image through learnable angles  $z_1^j, z_3^j$ and scalars $\delta_x^j, \delta_y^j$ respectively. We sample a discrete number of points along the ray, apply the shifts $\delta_x^j$ and $\delta_y^j$ along the $x$ and $y$ axes, and then perform a rotation $R(z_3^j)R(\theta^j) R(z_1^j)$, where $R$ are rotation matrices. We then evaluate the intensity at those points and integrate along the ray to find the predicted pixel intensity
$I_{\textrm{pred}}^{i,j}$.

The second term is a  total-variation regularization term, calculated as 
\begin{equation}
    \mathcal{L}_{\textrm{TV}} = \left <\left|\nabla_{\boldmath{r}}\phi(\boldmath{r})\right| \right>,
\end{equation}
where the average is calculated over a randomly sampled subset (due to memory constraints) of the coordinates evaluated to calculate the $\mathcal{L}_{\textrm{pixel}}$ term. The derivative can be calculated analytically through automatic differentiation.

We minimize the loss Eq.~(\ref{eq:loss}) with the Adam optimizer. We use a learning rate of $8\times10^{-6}$ for the neural-network weights, and a learning rate of $5\times10^{-2}$ for the pose parameters. We start with a warmup period for ten epochs where we gradually scale up the learning rate to the value of $8\times10^{-6}$, and then use a cosine-annealing learning rate scheduler to decrease it during training. To rapidly converge to a correct shape for the volume, we initially use very few sampling points along each ray and then ramp this up over the first few epochs. We use a batch size of 1024 pixels per GPU, resulting in a global batch size of 16384.

In scenarios where the signal-to-noise ratio in the tilt series images is high, such as at low electron fluence in Figs.~\ref{fig:phantom_results}, \ref{fig:experimental_results}, we divide the data set in a training set, which contains 80$\%$ of pixels in the data set (randomly sampled) and validation set which contains the other $20\%$. To prevent overfitting to noise, we evaluate the  $\mathcal{L}_{\textrm{pixel}}$ loss term on the validation set after every training epoch, and stop training once the loss on the validation set starts to increase. 

Training is straightforward to distribute over many GPUs by having a model replica on each GPU evaluate its own batch, and then averaging the calculated gradients (``Distributed Data Parallel''\cite{li2020pytorch}). As this results in a larger effective batch size, we can take larger gradients steps; we here scale the learning rate by the square root of the number of GPUs \cite{krizhevsky_one_2014,hoffer_train_2018}.

\subsection{Comparison with SIRT}

The SIRT algorithm used throughout this paper uses a vectorized \verb|PyTorch| re-implementation of the Radon and inverse-Radon transform from \verb|scikit-image| \cite{van_der_walt_scikit-image_2014}. To create the best possible reconstruction, we also implement a cross-correlation inline alignment of the tilt stack. In all SIRT reconstructions on this paper we also applied the positivity constraint on the volume.

\subsection{Electron fluence simulations}
For simulating electron fluence, we use the formula,

\begin{equation}
    \textrm{im}_\textrm{noisy} = \frac{\textrm{Poisson}(\textrm{im}\times \textrm{dose})}{\textrm{dose}}
\end{equation}

\noindent where we retrieve the image from the Poisson distribution.

\subsection{Phantom data set}

The phantom shown in Fig.~\ref{fig:phantom_results} is a $200\times200\times200$ volume that contains low- and high- spatial frequencies. In contrast to well-known phantoms such as the Shepp-Logan phantom, we find that it does not represent what a true materials system would consist of. In our phantom, we have a \textit{sea-mine}-like structure to simulate a similar structure to the TiO$_2$ with AuPd system we reconstructed. Including high-spatial frequency features such as a pyramid-like structure, and with atomic-like structures. Included is also a \textit{cheese-wedge} like structure to make sure we can reconstruct low-spatial frequency features.

We use an $\alpha=1\times10^{-4}$ for the total variational loss, with a $\beta=0.07$ for the $\textrm{smooth}_{L_1}$ loss. Each reconstruction was trained to 150 epochs. The volume sshown in Fig.~\ref{fig:phantom_results} converged within 30 epochs with no noticeable change.

\subsection{Catalytic nanoparticles dataset}\label{sec:method_catalytic}

The catalytic TiO$_2$-AuPd system was collected using the Thermo Fisher Spectra 300 with the Fischione 2040 dual-axis tomography holder. AuPd/TiO$_{2}$ were synthesized following a co-deposition-precipitation protocol. 
Prior to the synthesis, the TiO2 powder was dried in air at 100°C to remove moisture. 
The dried TiO$_{2}$ was dispersed into the aqueous solution containing pre-dissolved HAuCl$_{4}$ 3H$_{2}$O and Pd(aca)$_{2}$ as the precursors, and urea as the reducing agent. 
The synthesized supported catalyst was collected after proceeding the reaction overnight at 85 °C. The collected catalyst was dried under vacuum and subsequently reduced in the H$_{2}$ atmosphere at 400 °C for 1 hour, followed by annealing in Ar. 
The prepared catalyst was deposited onto a Simpore two-slot SiN window (SN100H-A05L), mounted on a Fischione 2040 dual-axis tomography holder. 

Since the dose sensitivity of the sample was unknown, the tilt series was collected by interleaving two tilt series each with a 4$^\circ$ step which resulted with a tilt range from -70$^\circ$ to 70$^\circ$ with 2$^\circ$ degree increments. 
Each tilt series subset in turn has their own respective tilt axis to be optimized over.
The original resolution of the projections was at 2048$\times$2048 resolution with a sampling rate of 0.1 nm per pixel.
For the reconstructions performed, each projection was scaled down to 500$\times$500 using Fourier downsampling.
We also collected 0$^\circ$ and 90$^\circ$ degree scan pairs for drift correction \cite{ophus_correcting_2016}.

To produce the final processed tilt series we performed drift-correction using the 0$^\circ$-90$^\circ$ scan pairs, background subtraction, rough cross-correlation, invariant line, and tilt axis alignment. For methods used for drift correction refer to \cite{ophus_correcting_2016}. For background subtraction we use bivariate Bernstein polynomial fitting to estimate the background. We then use cross-correlation to align the tilt stack, then used center-of-mass alignment to align the stack to the invariant line. Aligning the tilt axis was performed by eye until a reasonable reconstruction was obtained.

The experimental reconstructions using our method and SIRT in Figures \ref{fig:experimental_sirtvsml} and \ref{fig:experimental_results} are $500\times500\times500$ volumes using the preprocessed tilt series. We use an $\alpha=3\times10^{-4}$ for the total variational loss, with a $\beta=0.07$ for the $\textrm{smooth}_{L_1}$ loss. Each reconstruction was trained to 150 epochs.

\subsection{Hyperbranched nanoparticle dataset}

The prealigned dataset of the hyperbranched Co$_2$P nanoparticle was taken from Levin et al.~\cite{levin_nanomaterial_2016}. The only additional preprocessing step performed was background subtraction using bivariate Bernstein polynomial fitting similar to Sec.~\ref{sec:method_catalytic}. We use an $\alpha=1\times10^{-5}$ for the total variational loss, with a $\beta=0.07$ for the $\textrm{smooth}_{L_1}$ loss. Each reconstruction was trained to 150 epochs.

\section*{Acknowledgments}

This research used resources of the National Energy Research Scientific Computing Center, a DOE Office of Science User Facility supported by the Office of Science of the U.S. Department of Energy under Contract No. DE-AC02-05CH11231 using AI4Sci@NERSC NERSC award NERSC DDR-ERCAP0038157. CL, ARCM and CO were supported by Stanford University. The catalytic nanoparticle tilt series dataset was collected at the Stanford Shared Nano Facilities (SNSF). We also thank Lin Yuan for developing the sample preparation of the catalytic nanoparticles, and Stephanie Ribet for helpful information for acquiring tomographic datasets.

\section*{Data Availability}

Link to be added at submission

\section*{Code Availability}

The Python code is available in the in the open-source \href{https://github.com/electronmicroscopy/quantem/tree/tomography}{Quantitative Electron Microscopy (quantEM)} package. The current code is in its development phase, and will be fully implemented into \verb|quantEM| along with tutorial notebooks which a link will be added at submission.

%TC:endignore

\bibliography{sn-bibliography}

\newpage
\setcounter{figure}{0}
\setcounter{section}{1}
\setcounter{subsection}{0}
\renewcommand{\figurename}{Supplemental Figure}

\section*{Supplemental Information}

\subsection*{XEDS Measurement of Supported Nanoparticles}\label{supp_sec:eds}
\begin{figure}[htb]
    \centering
    \includegraphics[width=1\linewidth]{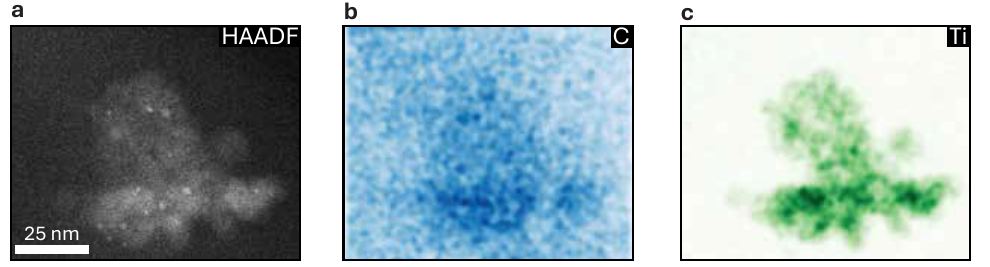}
    \caption{\textbf{HAADF, and Carbon and Titanium EDS maps of supported nanoparticles}. \textbf{(a)} HAADF image of the sample \textbf{(b)} Carbon \textbf{(c)} and Titanium XEDS signals.}
    \label{fig:catalytic_eds}
\end{figure}

To address the bimodality seen in Fig.~\ref{fig:experimental_results},we performed EDS measurements on a different sample found on the same TEM grid shown in Supp. Fig. ~\ref{fig:catalytic_eds}. Subsequent tomography experiments on the same sample were attempted by performing plasma cleaning of the grid prior to depositing the sample. Significant charging was observed, and tomography experiments were largely unsuccessful. We suspect that the presence of carbon proved vital for setting the nanoparticles and dramatically preventing charging of the sample. In Supp. Fig.~\ref{fig:catalytic_eds}b, there is non-neglibile signal at the base of the particle imaged, and in Supp. Fig~\ref{fig:catalytic_eds}c the titanium oxide support is clearly shown.

\subsection*{smooth$_{L_1}$ and $L_2$ Reconstruction Comparisons}\label{supp_sec:l1l2comp}

\begin{figure}[htb]
    \centering
    \includegraphics[width=1\linewidth]{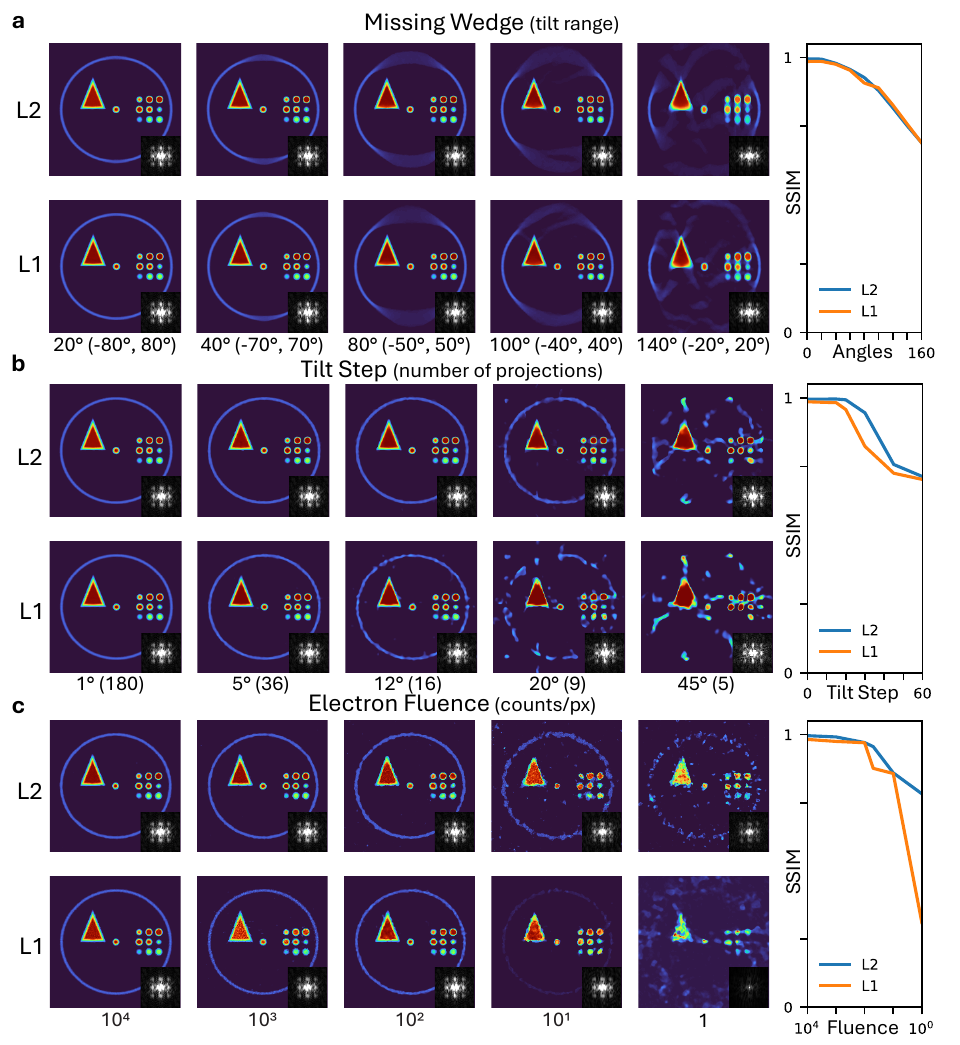}
    \caption{\textbf{Comparison between smooth$_{L_1}$ (L1) and MSE (L2) loss for a simulated phantom}. \textbf{a} \textbf{(a-c)} Comparison between L1 and L2 losses by varying the missing wedge size, tilt step, and electron fluence respectively through a slice perpedincular to the missing wedge direction. The SSIMs were computed using the ground truth slice.}
    \label{fig:phantom_l1l2}
\end{figure}

The choice of loss function has a significant effect on the quality of the reconstruction. To see the effect of the two loss functions we use the simulated phantom using a TV weight of $\alpha=10^{-4}$, and for smooth$_{L_1}$ (L1) using a value of $\beta=0.07$ as described in the main text. In Supp. Fig.~\ref{fig:phantom_l1l2}a, no noticeable change can be seen between the two losses. This can be further seen in the SSIM where the two values are similar. Varying the tilt step in Supp. Fig.~\ref{fig:phantom_l1l2}b for the $12^\circ$ case, L2 is able to more effectively retain the full structure of the membrane. In contrast, the L1 loss starts to lose continuity across the membrane while still retaining the internal features. The SSIM obtained with L1 is marginally less robust to larger tilt steps compared to L2, while converging to the same performance at very limited projection cases. The most significant change can be seen when varying the dose in Supp. Fig.~\ref{fig:phantom_l1l2}c. 
At low dose cases, in particular the 10 counts/px, L1 significantly dims the membrane while L2 is still able to capture the membrane. 
In the 1 counts/px case, L2 significantly outperforms L1 in retaining the membrane which leads to a large difference in the SSIM.

\begin{figure}[htb]
    \centering
    \includegraphics[width=1\linewidth]{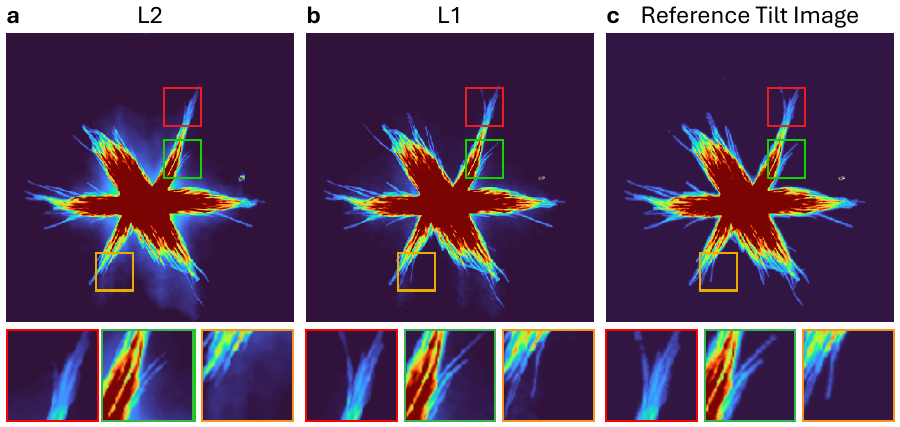}
    \caption{\textbf{Comparison between smooth$_{L_1}$ (L1) and MSE (L2) loss on the hyperbranched Co$_2$P nanoparticle.} Reconstructions of the particle using L2 \textbf{(a)}, L1 \textbf{(b)}, and the reference tilt image \textbf{(c)}. Zoomed-in regions are shown highlighting missing features.} 
    \label{fig:nanostar_loss_comp}
\end{figure}

While the results discussed above would suggest that using an L2 pixel-wise loss would significantly improve reconstruction quality, applying this to the hyperbranched nanoparticle shows otherwise. For this case we set the TV weight $\alpha=0$ and $\beta=0.07$; relying solely on the pixelwise losses. In Supp. Fig.~\ref{fig:nanostar_loss_comp}, we compare the ground truth reference tilt image with the corresponding summed down projection (X-Y plane) of the L2 and L1 reconstructions. Comparing Supp. Fig.~\ref{fig:nanostar_loss_comp}a to Supp. Fig.~\ref{fig:nanostar_loss_comp}c, real features are not present or significantly reduced. In contrast to the L1 case in Supp. Fig.~\ref{fig:nanostar_loss_comp}b, the features in the reference tilt image is effectively being reproduced.

The results shown here may give some insights on how to direct hyperparameter optimization of $\beta$ and $\alpha$. In particular, for samples that are dominated by high spatial frequencies requires lower $\beta$ and $\alpha$ values which effectively turns the pixelwise loss function to a pure $\mathcal{L}_1$ loss. For samples dominated by low spatial frequencies, using a higher $\beta$ may offer better performance. However, fine tuning of $\alpha$ may still be necessary.

\end{document}